\documentclass{article}

\usepackage{graphicx} 
\usepackage{subfigure} 

\usepackage{natbib}

\usepackage{algorithm}
\usepackage{algorithmic}

\usepackage{amsmath}
\usepackage{amsfonts}

\usepackage{hyperref}

\usepackage[accepted]{icml2012}

\icmltitlerunning{Finding cancer subtypes in glioblastoma}
\begin{document} 
\twocolumn[
\icmltitle{Identifying cancer subtypes in glioblastoma by combining
  genomic, transcriptomic and epigenomic data}

\icmlauthor{Richard S. Savage}{r.s.savage@warwick.ac.uk}\icmladdress{Systems Biology Centre, University of Warwick, Coventry, CV4 7AL, UK}
\icmlauthor{Zoubin Ghahramani}{} \icmladdress{Department of Engineering,University of Cambridge,  Cambridge, CB2 1PZ, UK}
\icmlauthor{Jim E. Griffin}{} \icmladdress{School of Mathematics, Statistics and Actuarial Science, University of Kent, CT2 7NF, UK}
\icmlauthor{Paul Kirk}{}\icmladdress{Systems Biology Centre, University of Warwick, Coventry, CV4 7AL, UK}
\icmlauthor{David L. Wild}{}\icmladdress{Systems Biology Centre, University of Warwick, Coventry, CV4 7AL, UK}

\icmlkeywords{cancer, glioblastoma, data fusion, data integration,
  Bayesian, nonparametric, machine learning, ICML}

\vskip 0.3in
]
\begin{abstract} 
We present a nonparametric Bayesian method for disease subtype
discovery in multi-dimensional cancer data.  Our method can
simultaneously analyse a wide range of data types, allowing for both
agreement and disagreement between their underlying clustering
structure.  It includes feature selection and infers the most
likely number of disease subtypes, given the data.  

We apply the method to 277 glioblastoma samples from The Cancer
Genome Atlas, for which  there are gene expression, copy number
variation, methylation and microRNA data. We identify 8 distinct
consensus subtypes and study their prognostic value for death, new tumour
events, progression and recurrence.  
The consensus subtypes are prognostic of tumour
recurrence (log-rank p-value of $3.6 \times 10^{-4}$ after correction for
multiple hypothesis tests).  This is driven principally by the
methylation data (log-rank p-value of $2.0 \times 10^{-3}$) but the
effect is strengthened by the other 3 data types, demonstrating the
value of integrating multiple data types.

Of particular note is a subtype of 47 patients characterised by very
low levels of methylation.  This subtype has very low rates of tumour
recurrence and no new events  in 10 years of follow up.  We also
identify a small gene expression subtype of 6 patients that shows
particularly poor survival outcomes.  Additionally, we note a
consensus subtype that showly a highly distinctive data  signature and
suggest that it is therefore a biologically distinct subtype of glioblastoma.

We note that while the consensus subtypes are highly informative,
there is only partial overlap between the different data types.  This
suggests that when considering multi-dimensional cancer data, the
underlying biology is more complex than a straightforward set of
well-defined subtypes.  We suggest that this may be a key
consideration when modeling such data.

The code is available from https://sites.google.com/site/multipledatafusion/

\end{abstract} 
\section{Introduction}

Cancer is a complex disease,  driven by a range of genetic and
environmental  effects.  It is responsible for 1 in 8 deaths worldwide
\citep{globalCancerStats}, with an estimated 7.6 million cancer deaths
worldwide in 2008 \citep{jemal2011global}. 
Understanding the cancer genome \citep{stratton2009} and the
associated molecular mechanisms is therefore a vitally important
global medical issue.

Modern large-scale cancer studies present great new opportunities to
understand different types of cancer and their underlying mechanisms.
Projects such as The Cancer Genome Atlas (TCGA) \citep{tcga} and METABRIC
\citep{metabric} and the International Cancer
Genome Consortium \citep{icgc}, are producing large, multi-dimensional
data sets that have the potential to revolutionise the study of cancer.

One of the first TCGA projects is a study of glioblastoma,
\citep{glioblastomaData} the most common primary brain tumour in human
adults.  Glioblastoma is an aggressive cancer; patients with newly
diagnosed glioblastoma have a median survival of $\approx 1$ year.
The TCGA glioblastoma data set is a hugely relevant resource for improving this situation.

Key to the utilisation of these multi-dimensional data sets is to
develop effective data fusion methods  
\citep[see e.g.][]{Shen2009, Savage2010, psdf, mdi}.
It is not enough to simply
concatenate the different data types; one must account for the
different statistical characteristics of each data type and that they
may contain differing or even contradictory information about the
samples studied.
There is potentially huge benefit
to proper analysis of such multi-dimensional data sets (e.g. consider
the number of pairwise data type comparisons as the total number of
types increases).  But to fully realise this, we must develop new methods.

The Multiple Data Integration (MDI) algorithm
\citep{mdi} is a principled framework for the identification of cancer
subtypes.  It can analyse \emph{multi-dimensional} data sets,
combining a range of individual data types such as gene expression,
copy number variation, methylation and microRNA data.  MDI can be
regarded as the extension to multiple (possibly disagreeing) data types of
nonparametric Bayesian clustering methods such as the Dirichlet Process
(DP) mixture model \citep[see e.g.][]{Ferg, Anton, EscWes:94, Dahl, Rasmussen}.

The key advantage of MDI is that it allows for the possibility of
both agreement and disagreement between the clustering structures of
different data types within a given analysis.  This is extremely
important in biological data sets.   For example, gene expression can be
regulated by a number of biological mechanisms, so is not determined
solely by the underlying genome.  Hence integrating gene expression
and copy number variation data might or might not result in good
agreement, depending on the biological context.

MDI produces clustering partitions for each data type, as well
as an overall \emph{consensus} clustering partition.  It also
identifies the degree to which different data types share common
structure, and can identify which of the items are fused across the
different data types.  

To extend the MDI method to the analysis of cancer data, we have added
additional functionality beyond that of \citet{mdi}  Two data models (Gaussian and
Multinomial) are used.  For each of these data models, feature
selection has been added, so that the most informative features can be
identified for each data type.  Additionally, it is known that
the MCMC chains in mixture-based clustering methods can be slow to mix
when using a Gibbs sampler.  To improve performance in this regard, an
additional split-merge MCMC sampler has been added, which is used in
conjunction with the existing Gibbs steps.  

MDI therefore has the following advantage in analysing post-genomic
molecular cancer data.
\begin{itemize}
\item{Infers (Rather than assumes) the degree to which clustering
    structure is shared  between data types}
\item{Infers the likely number of clusters given the data}
\item{Identifies the genes/probes in each data type that define the
    disease subtypes}
\item{Integrate simultaneously  a wide range of data
    types (4 in this paper; more can easily be included if available)}
\end{itemize}

The rest of the paper is summarised as follows.  In Section \ref{data}
we describe the data set.  In Section \ref{model} we
describe MDI and present several improvements to the method.  In
Section \ref{results} we present the results of analysing the TCGA
glioblastoma data.  Finally, in Section \ref{conclusions} we draw
conclusions about this work.

\section{Data}\label{data}
We downloaded glioblastoma data \citep{glioblastomaData} 
from the TCGA data portal (\emph{http://cancergenome.nih.gov/}), including gene expression,
copy number variation, methylation and microRNA data, as well as
clinical follow-up information.
After matching samples across all 4 data types, we are left with  277
samples for which we have complete data.
We note that in a few cases (and for a given data type) there are
duplicate samples for the same patient.  In this case we make a blind
selection of the first sample, based on bar code ordering.

All data were downloaded from the TCGA data portal on 13th April 2012.

\subsection{Gene Expression}
We use the publicly-available level 3 gene expression data.
For consistency, data were chosen from a single platform.  The UNC
AgilentG4502A 07 samples were selected as they were most numerous,
giving a total of 571 tumour samples.
These were read into a single data matrix and NaN (missing) values set
to zero.  

The data include 10 normal samples.  For each gene, a
Wilcoxon rank-sum test was used to determine whether or not there was
differential expression between tumour and normal  samples.  A
Bonferroni correction and p-value threshold was applied ($p<2\times
10^{-3}$), leaving 1011 gene expression features.

\subsection{Copy Number Variation}
We use the publicly-available level 2 copy number data.  We chose
level 2 so that we had access to all probes (the level 3 data are
segmented into regions which, in general, are different from sample to
sample). This gave 466 tumour and 376 normal samples generated by MSK
C using the HG-CGH-244A platform.   These were read into a single data
matrix and NaN (missing) values set to zero.  

For each probe, a Wilcoxon rank-sum test was used to determine
whether or not there was differential copy number variation between
the normal and tumour samples.  A
Bonferroni correction was then applied to the p-values.
A large number of probes had highly significant p-values, many of
which contained similar information.  It was therefore decided on
practical grounds to keep only the 1000 most significant probes as
features for this analysis.

\subsection{Methylation}
We use the publicly-available level 3 methylation data.
This gave 285 samples generated by  JHU-USC on the HumanMethylation27
platform.
The data were in the form of \emph{beta values}, which measure the
ratio of methylation signal to (methylation + background) signal.  For
convenience, the data were binarised using a threshold of $\beta >
0.95$.  Features containing fewer than 10 hits were then removed,
leaving 769 features.

\subsection{Micro RNA}
We use the publicly-available level 3 microRNA data.
This gave 490 tumour samples generated by UNC on the H-miRNA\_8x15K platform.

The data include 10 normal samples.  For each microRNA in turn, a
Wilcoxon rank-sum test was used to determine whether or not there was
differential expression between tumour and normal samples.  Applying a Bonferroni correction and then
keeping only genes with a p-value of $p<1 \times 10^{-3}$ gave us 104 features. 

\subsection{Clinical Follow-up}
The corresponding clinical data were also downloaded.  The files
follow\_up\_v1.0\_public\_GBM.txt and 
clinical\_patient\_public\_GBM.txt 
were used, matching the patients on the basis of the TCGA bar codes.
We note that 51 of the 277 samples did not have complete clinical follow-up
information.  

\begin{figure}[!ht]\centering
\includegraphics[width=.75\linewidth]{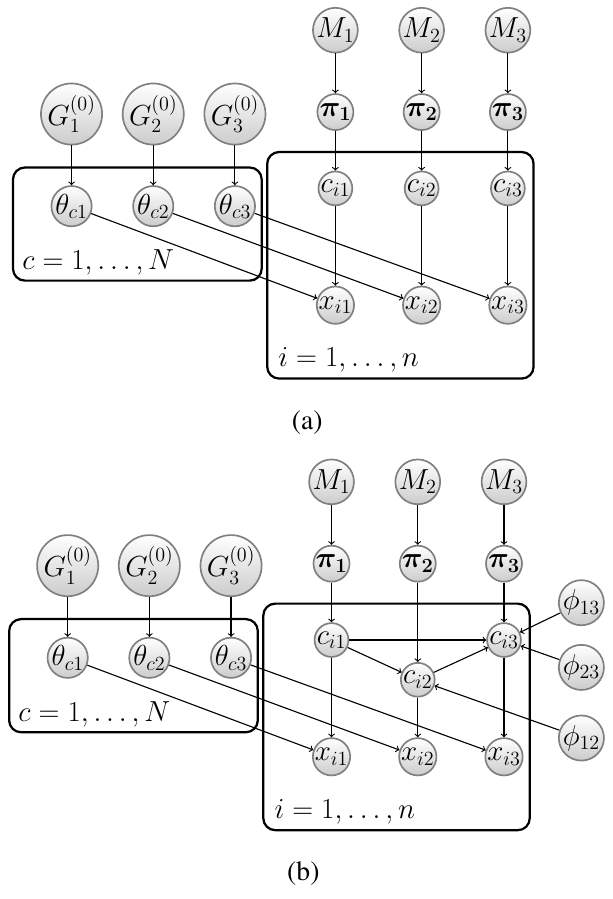}
\vspace{-2ex}
  \caption{Graphical representation of $K=3$ DMA mixture models.  (a)
    Independent case.  (b) Modelling dependence between the latent
    component allocation variables (the MDI model).  Figure from \citet{mdi}}
\label{graphs}
\end{figure}  

\section{Model}\label{model}
We further develop the Multiple Data Integration (MDI)
method of \citet{mdi}.
MDI can be regarded as an extension of nonparametric Bayesian
clustering methods to analyse simultaneously multiple data types,
inferring the degree of similarity between the clustering structure in
the data types.
MDI hence produces clustering partitions for each data type, as well
as an overall \emph{consensus} clustering partition.  

We note that another particular strength of MDI is that it infers
the posterior distribution over the number of clusters in each data
type.  Hence, we can infer the likely number of clusters, given the
data.  Many clustering algorithms do not provide a method for doing
this, which is a major shortcoming.

\subsection{The MDI model}
We consider a multi-dimensional data set, consisting of $K$ distinct
data types (for example, gene expression, copy number, methylation or
microRNA expression).  Each data type will contain measurements for the same set
of items, so that for each items we have $K$ vectors of measurements.
We note that in general the numbers of features for each data type
will be different from one another, and can be arbitrarily so for the
MDI model.

We model each data type using a finite approximation to a Dirichlet
process mixture model \citep{Ishwaran2002}, known as a
Dirichlet-Multinomial Allocation (DMA) mixture model \citep{Green2001}.
The $K$ DMA models are coupled by a set of $\phi$ parameters that
allow for information-sharing between the data types and provide a
measurement of the level of similarity between pairs of types.
Figure \ref{graphs} shows a graphical representation of the MDI model.

The DMA mixture model for a single data type is given by the following
equation.
\begin{equation}
p(x) = \sum_{c = 1}^N \pi_c f(x|\theta_c).\label{1}
\end{equation}
Where $\pi_c$ are the mixture proportions, $f$ is a parametric
density (such as a Gaussian) that models the $c$-th data type,  
and $\theta_c$ are the parameters associated with the $c$-th
component.  
$N$ is the maximum number of mixture components, which is 
set typically  to be large enough so as to not impact on the
inference.  When $N$ is large in this way, the behaviour of the DMA
model approaches that of a Dirichlet process.  To ensure this and as 
a tradeoff with computational cost, $N = \lceil n/2 \rceil$ throughout
this paper.  

We note that different choices for the density $f$  allow
us to model different types of data (for example, a normal
distribution might be appropriate for continuous data, while a
multinomial might be appropriate for categorical data).  This imparts
great flexibility to the MDI model.
 
Given observed data $x_1, \ldots, x_n$, we wish to perform Bayesian
inference for the unknown parameters in this model.  We introduce
latent {\em component allocation} variables $c_j \in \{1, \ldots,
N\}$, such that $c_i$ is the component responsible for observation
$x_i$.  We then specify the model as follows:  
\begin{align}
x_i | c_i, \boldsymbol{\theta} &\sim F(\theta_{c_i}),\notag\\
c_i |\boldsymbol{\pi}&\sim \mbox{Multinomial}(\pi_1, \ldots, \pi_N),\notag\\
\pi_1, \ldots, \pi_N &\sim \mbox{Dirichlet}(\alpha/N, \ldots, \alpha/N),\label{2}\\
\theta_c &\sim G^{(0)}\notag,
\end{align}
where $F$ is the distribution corresponding to density $f$,
$\boldsymbol{\pi}= (\pi_1, \ldots, \pi_N)$ is the collection of $N$
mixture proportions, $\alpha$ is a mass/concentration parameter (which
may also be inferred), and $G^{(0)}$ is the prior for the component
parameters. 

Now considering the full MDI model with $K$ distinct data types, to
couple the $K$ mixture models, the following conditional prior is
introduced for the component allocation variables.
\begin{equation}\label{generalModel}
p(c_{i1},\dots,c_{iK} | \boldsymbol{\phi})\propto \prod_{k=1}^K \pi_{c_{ik}k}
\prod_{k=1}^{K-1}\prod_{\ell=k+1}^{K}
\left(1 + \phi_{k \ell}{\mathbb{I}(c_{ik}=c_{i\ell})}\right),
\end{equation}
where $\mathbb{I}$ is the indicator function, $\phi_{k \ell} \in
\mathbb{R}_{\ge 0}$ is a parameter that controls the strength of
association between datasets $k$ and $\ell$, and $\boldsymbol{\phi}$
is the collection of all $K(K-1)/2$ of the $\phi_{k \ell}$'s. 
$c_{ik} \in \{1, \ldots, N\}$ is the component allocation variable
associated with item $i$ in model $k$, and  
$\pi_{c_{ik}k}$ is the mixture proportion associated with component
$c_{ik}$ in model $k$.  
Informally, the larger $\phi_{k\ell}$, the more likely it is that
$c_{ik}$ and $c_{i\ell}$ will take the same value, and
hence the greater the degree of similarity between the clustering
structure of dataset $k$ and dataset $\ell$.  
If all $\phi_{k\ell} = 0$, then we recover the case of $K$ independent DMA
mixture models (Figure~1b).  We constrain the $\phi_{k \ell}$ to be
non-negative.

We note that the $x_{ij}$ are assumed to be independent, given the
clustering in MDI.  The model then concentrates on modeling the joint
distribution of the allocation variables $c_{i1},...,c_{iK}$ which
induces correlation over the x's.

\subsection{Data Models}
For the analyses in this paper, we use two different densities,
$f_{Gaussian}$ and $f_{multinomial}$.  These are respectively used for
real-valued and discrete data and make reasonable assumptions (for the
data used in this paper) about the expected noise characteristics.  

For both data models, we assume that the features represent repeated,
independent measurements of the underlying clustering partition.  We
therefore have the following equation.
\begin{equation}\label{features}
f = \prod_a f_a
\end{equation}
For the Gaussian density, we assume that each feature is modeled by a
Gaussian likelihood of unknown mean and precision, subject to a
Normal-Gamma prior.  There density is therefore closed form and we
hence marginalise the mean and precision.

We therefore have the following (marginal) likelihood function for the
Gaussian case.
\begin{equation}
f_{a, Gaussian} = \frac{\Gamma(\alpha_n)}{\Gamma(\alpha_0)}
                      \frac{\beta_0^{\alpha_0}}{\beta_n^{\alpha_n}}
                      (\frac{\kappa_0}{\kappa_n} )^\frac{1}{2}
                      (2 \pi)^{-\frac{n}{2}}
\end{equation}
Where
\begin{eqnarray}
\mu         &\sim& N(0, (\kappa_0 \lambda)^{-1})\\
\lambda  &\sim& Ga(\alpha_0, \beta_0)\\ 
\kappa_n &=&      \kappa_0 + n\\
\alpha_n  &=&      \alpha_0 + \frac{n}{2}\\
\beta_n    &=&      \beta_0 + \frac{1}{2}\sum_{i=1}^n(x-\bar{x})^2 +
\frac{\kappa_0 n \bar{x}^2}{2 \kappa_n}
\end{eqnarray}
We set the Normal-Gamma hyperparameters to $\alpha_0=2$, $\beta_0=0.5$
and $\kappa_0=0.001$.

For the multinomial density, we assume that each feature is modeled
by a multinomial likelihood, subject to a Dirichlet prior.  The
parameters of the multinomial likelihood are unknown, but because of
the conjugate prior the density has a closed form and those parameters
can hence be marginalised, leaving only the hyperparameters
$\beta_{rq}$ to be defined.

We therefore have the following (marginal) likelihood function for the
multinomial case.
\begin{equation}
f_{a, multinomial} = \prod_{q=1}^Q
\frac{\Gamma(B_q)}{\Gamma(S_q+B_q)}\prod_{r=1}^R \frac{\Gamma(x_{rq} +
  \beta_{rq})}{\Gamma(\beta_{rq})}
\end{equation}
 We set the Dirichlet prior hyperparameters to $\beta_{rq} = 0.5$.

\subsection{Feature Selection}
Because of the potentially large number of features in the various
omic data types, we extend the Gaussian and multinomial data models to
include feature selection.  To do this, we include binary indicator
parameters $I_a$ for each feature in a given data type.  This modifies
Equation \ref{features} to the following:

\begin{equation}
f = \prod_a (I_a f_a) + (1-I_a)f_{a, null}
\end{equation}

The factor for each feature will therefore either be $f_a$ or $f_{a, null}$.
The $f_a$ are as before, so if all $I_a = 1$ then we have the model
with no feature selection.  

The $f_{a, null}$ represent the alternative model that the feature is
uninformative and hence all items are modeled as belonging to a single
mixture component.  We also make an approximation that the likelihood
parameters are known, rather than marginalised over.  This makes only
a modest correction to the typical marginal likelihood values, but
significantly speeds up the computation of the conditional
distributions for Gibbs resampling.

For data models taking the form in Equation \ref{features}, we note
that the distributions for Gibbs resampling of the
$I_a$ are conditionally independent, given the $c_{ik}$.  As MDI is
written in Matlab, this allows us to vectorise the computation of the
conditional distributions for the Gibbs resampling of all the $I_a$.
This vectorisation makes the feature selection in MDI highly
computationally efficient and fast to execute.

\subsection{Split-Merge MCMC sampling}
One characteristic of Gibbs samplers for mixture model clustering
algorithms is that the MCMC chains can be relatively slow to mix.  We
have noticed this in particular for the number of occupied components.  

To improve this, we have implemented a version of the sequential
split-merge MCMC sampler of \citet{dahl2005}.  The split-merge steps
are applied separately to each of the $K$ DMA models.  
These MCMC steps are proposed  in addition to all the usual Gibbs
steps described in \citet{mdi}.  The increase in computation required
for the split-merge steps is minor, and we find that while the
acceptance rate for the steps is low, the overall effect is to
substantially improve the mixing rate for the number of clusters for
each data type.

\subsection{Extraction of Clustering Partitions}
We adopt a different approach to that of \citet{mdi} to the
extraction of clustering partitions from the posterior similarity
matrix.  Because the previously-used method \citep{Fritsch2009} is
only available as an R package, for convenience we implement as part
of the MDI code a simpler method based on a hierarchical
clustering using the posterior similarity matrix. We note that the
results in \citet{Fritsch2009} show that this approach produces similar
performance to the previously-used method.

Using as distances (1 - posterior similarity), we perform standard
hierarchical clustering with complete linkage.  We set the number of
clusters to be the MAP estimate of the number of clusters, taken from
the MCMC analysis.

The resulting clustering partition should be regarded as a convenient
summary of the full results.  We would strongly encourage users of
this method to not neglect other outputs such as the posterior
similarity and fusion matrices.

\begin{table*}[t]
  \caption{Bonferroni-corrected p-values for subtype Kaplan-Meier
    survival curves (nTests=20) For a given clinical outcome and data
    type/s, the Kaplan-Meier curves for each disease subtype are
    produced.  P-values are computed using the log-rank test and
    considering the null hypothesis that all curves in a given set are
    drawn from the same underlying distribution.
 } 
  \begin{center}  
   \begin{tabular}{| l | c | c | c | c |}
      \hline
      \small
      data type/s &\small died &\small new event &\small progression &\small recurrence\\
      \hline
      \small all                      & 1& 0.66& 0.11& $3.6 \times 10^{-4}$\\
      \small copy number     & 1& 1     & 1     &1\\
      \small gene expression& $1.5 \times 10^{-3}$& 1& 1& 1\\
      \small methylation       & 1& 0.28& 1& $2.0 \times 10^{-3}$\\
      \small microRNA          & 1& 1& 1& 1\\
     \hline
    \end{tabular}
  \end{center}
  \label{Table:qvalues} 
\end{table*}

\section{Results}\label{results}
We analyse gene expression, copy number variation, micro-RNA and
methylation data for 277 glioblastoma patients using MDI.  We identify
a range of distinct disease subtypes and results, the most interesting
of which we now describe.

Complete results can be found at https://sites.google.com/site/multipledatafusion/

We consider 5 distinct cases of summarising clustering partition (for
each individual type, and also the consensus of all 4).  For these we
consider 4 binary, right-censored clinical outcomes: death, new tumour
event, tumour progression, tumour recurrence.  This gives us 20 cases in all. 

For each case, we plot the Kaplan-Meier survival curves for the
set of disease subtypes.  Table \ref{Table:qvalues} shows the log-rank
p-values for  these plots, after application of a Bonferroni
correction for multiple hypothesis testing (nTests=20).
Examples of these plots can be seen in Figure \ref{fig:survival_recurrence}.
We note that in all cases we only consider subtypes containing at
least 5 items and we only consider items for which there is clinical
outcome information.

\subsection{Consensus subtypes are prognostic for tumour recurrence}
We note that the consensus subtypes in general are strongly
prognostic for tumour recurrence  (log-rank p-value of $3.6 \times
10^{-4}$ after correction for multiple hypothesis tests) (see Figure \ref{fig:allSortedData}).  Because the
methylation status is measured at the point of diagnosis, this
prognostic capability is predictive.

\subsection{Interesting low-methylation subtype}
Consensus subtype 7 has particularly interesting characteristics.
Comprising 47 items, it shows only very low levels of tumour
recurrence (see Figure \ref{fig:survival_recurrence}) and no new
tumour events.  All items in this
subtype show very low relative levels of methylation (see Figure
\ref{fig:methylationLevels}). 
This subtype contains 23 women and 24 men, with a median age of 53 and
an age range of 14 to 81.

\subsection{Gene expression subtypes are prognostic for survival outcome}
Figure \ref{fig:survival_death} shows the Kaplan-Meier survival curves
for the 8 gene expression (GE) subtypes identified by MDI.  These subtypes
are prognostic for survival outcome (log-rank p-value of $1.5 \times
10^{-3}$ after correction for multiple hypothesis tests).

This result is largely driven by GE cluster 7, which consists of 6
patients with particularly poor survival outcome (see Figure
\ref{fig:survival_death}).  Of these 6 patients, 3 die within 6 months
of diagnosis, and the other 3 are omitted from the survival analysis
as we do not have information on their survival or otherwise (they are
missing data).  As such, this result relies on a small number of
patients and should therefore be treated with caution.  However,
further study is certainly warranted in case this subtype remains
distinct with a larger number of members.

\subsection{Partial overlap of clustering structure for different data types}
The fusion matrix (Figure \ref{fig:fusionMatrix} ) and consensus
clustering results show that there is a level of consistency in the
clustering structure across the gene expression, copy number
variation, methylation and microRNA data.  However, inspection of the
clustering partitions for each data type show that there are also
differences in structure between each type.  This indicates that a
single clustering partition is not sufficient to capture all of the
structure contained in the 4 data types.

\subsection{Evidence for a biologically distinct glioblastoma subtype}
Consensus cluster 5 consists of 8 patients and is noteworthy for
highly distinctive data signatures in gene expression
(over-expression), copy number (excess copies) and micro-RNA
(over-expression in a subset of selected features).

This subtype is poor for tumour recurrence, and we suggest that the
striking data signatures are suggestive of a distinctive set of
biological mechanisms driving this tumour subtype.

\subsection{Fusion matrix makes biological sense}
\begin{figure}
\centering
\vspace{-22ex}
\includegraphics[width=1.\linewidth]{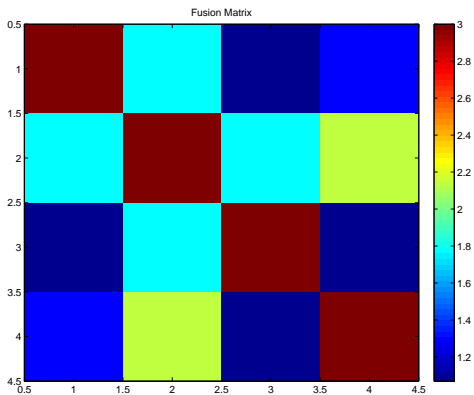}
\vspace{-22ex}
\caption{Matrix showing the posterior mean values for the
  $\phi_{k\ell}$.  Shown are the copy number data (row 1), gene
  expression (row 2), methylation (row 3), microRNA (row 4).  
  We note that the diagonal elements are undefined for the MDI model and
  so are set to arbitrary values so as to make the colour table convenient.}
\label{fig:fusionMatrix}
\end{figure}

The $\phi_{k\ell}$ parameters provide information on the level of
agreement about clustering structure between pairs of data types.
Figure \ref{fig:fusionMatrix} shows the posterior mean values for the
  $\phi_{k\ell}$.  
The principal sharing of structure is shown to be between gene
expression and the other three data types, while other pairs of data
types are less strongly related.  This confirms what would be
expected from prior knowledge of the underlying biological
mechanisms.
This is an important sanity check of the analysis, and shows that
there is useful biological information in the data set.

\subsection{MCMC details}
The results presented in this paper are the result of 25 MCMC chains,
each of $\approx 70,000$ samples.  We sparse-sample the chains by a
factor of 10 and remove the first 25\% of each chain as burn-in.  We
check for adequate convergence by visual inspection of the MCMC
time-series and histograms for each chain, overlaid on one another.

\section{Conclusions}\label{conclusions}
\begin{figure}
\centering
\includegraphics[width=1.\linewidth]{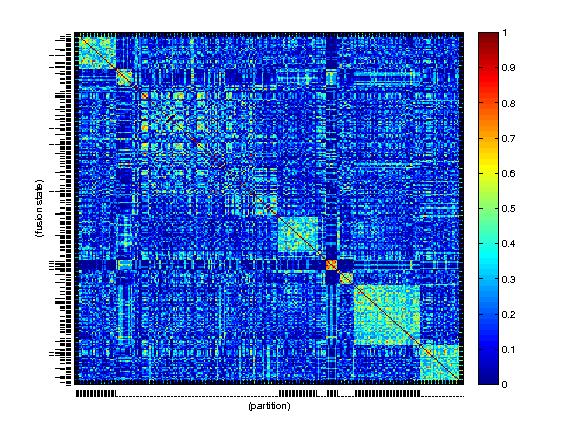}
\vspace{-6ex}
\caption{Posterior similarity matrix for the 277 data items.  This
  matrix gives the posterior probability of given pairs of items
  belonging to the same cluster, averaged over the 4 data types.  This
is used to produce the consensus clustering partition.  The matrix
shown here has been sorted by the resulting partition, to better show
the structure.  The x-axis is labelled by alternating sets of
bars/dots, denoting the different clusters.  The LHS y-axis bars show
the relative probability  that a given item has the same cluster label
across data types.}  
\label{fig:allSimilarityMatrix}
\end{figure}  

We have presented extensions to the MDI method that make it suitable
for analysing multi-dimensional molecular cancer data sets.
Using MDI to analyse the TCGA glioblastoma data, we have identified a
number of important points.
\begin{itemize}
\item{Both the 8 methylation and 8 data-consensus disease subtypes we
    have identified are significantly prognostic of tumour recurrence}
\item{The 8 gene expression disease subtypes we have identified are significantly
    prognostic of survlval outcome}
\item{We have identified a strongly prognostic glioblastoma subtype,
    noteworthy for its low levels of tumour recurrence methylation}
\item{We have identified a small subtype of 6 patients based on gene
    expression for which there is very poor survival outcome, and
    postulate that this may identify a rare and agressive form of glioblastoma}
\item{We have also identified a subtype of 8 patients with a highly distinctive data
  signature.  This subtype has poor tumour recurrence, and it may
  represent a subtype whose underlying biology is highly distinctive,
  which may allow for more targetted therapy}
\item{We note that the clustering structures for the 4 different data types overlap
  partially, but there is significantly more structure than can be
  explained by a single partition}
\end{itemize}

Modern, large-scale cancer data sets contain a wealth of data types
measuring the effects of genomic and environmental processes.   We have
demonstrated the effectiveness of combining these data types into a
single analysis, and shown that the richness of structure contained by
such multi-dimensional data sets necessitates statistical methods
capable of capturing that richness.

\begin{figure*}
\vspace{-15ex}
\includegraphics[width=0.9\linewidth]{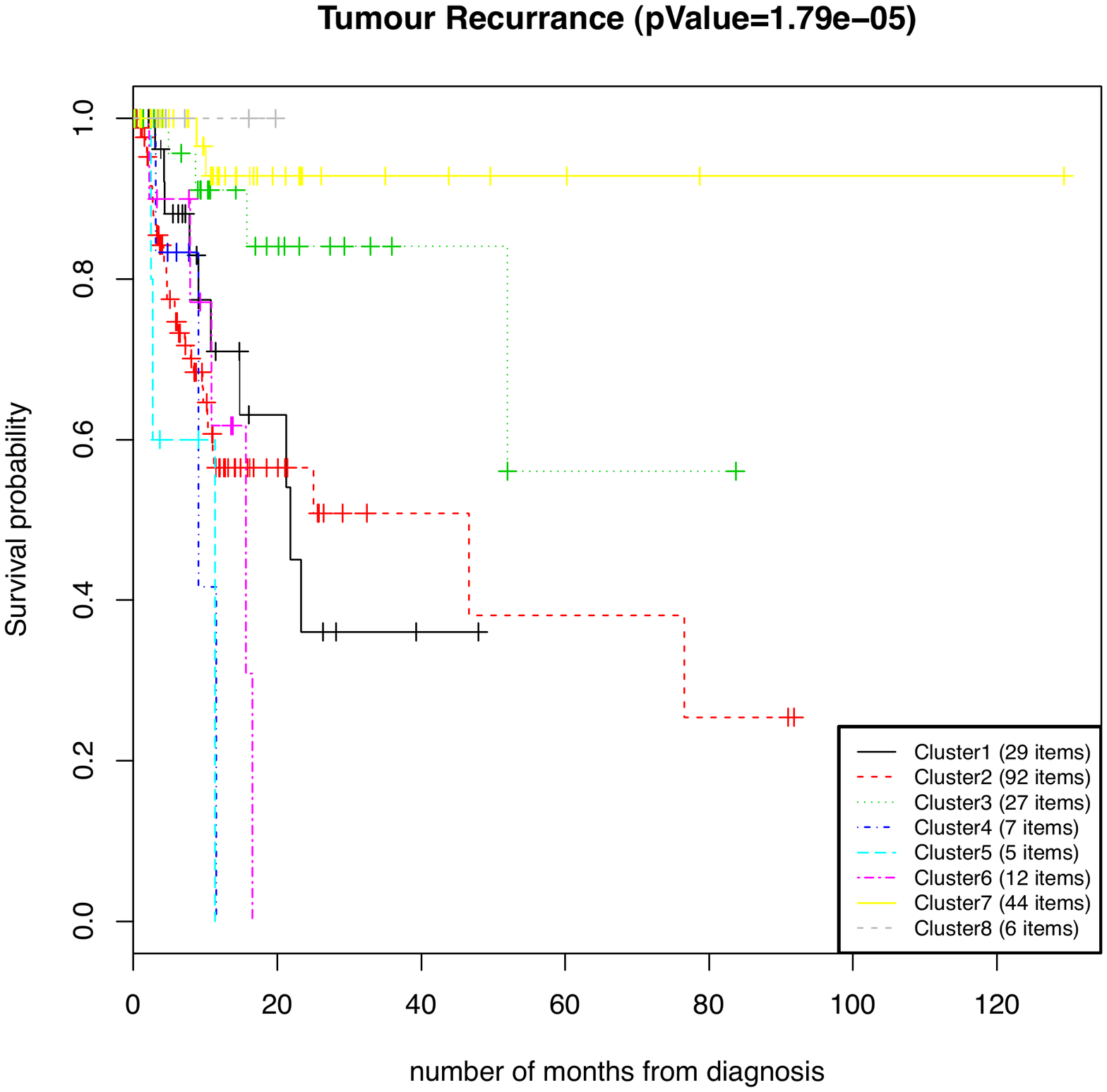}
\vspace{-15ex}
\caption{Kaplan-Meier survival curves plotting tumour recurrence for
  the consensus disease subtypes.  The p-value is computed using a
  log-rank test.  When quoted in Table \ref{Table:qvalues}, a
  Bonferroni correction has been applied to account for multiple
  hypothesis tests.}
\label{fig:survival_recurrence}
\end{figure*}  

\begin{figure*}
\vspace{-15ex}
\includegraphics[width=0.9\linewidth]{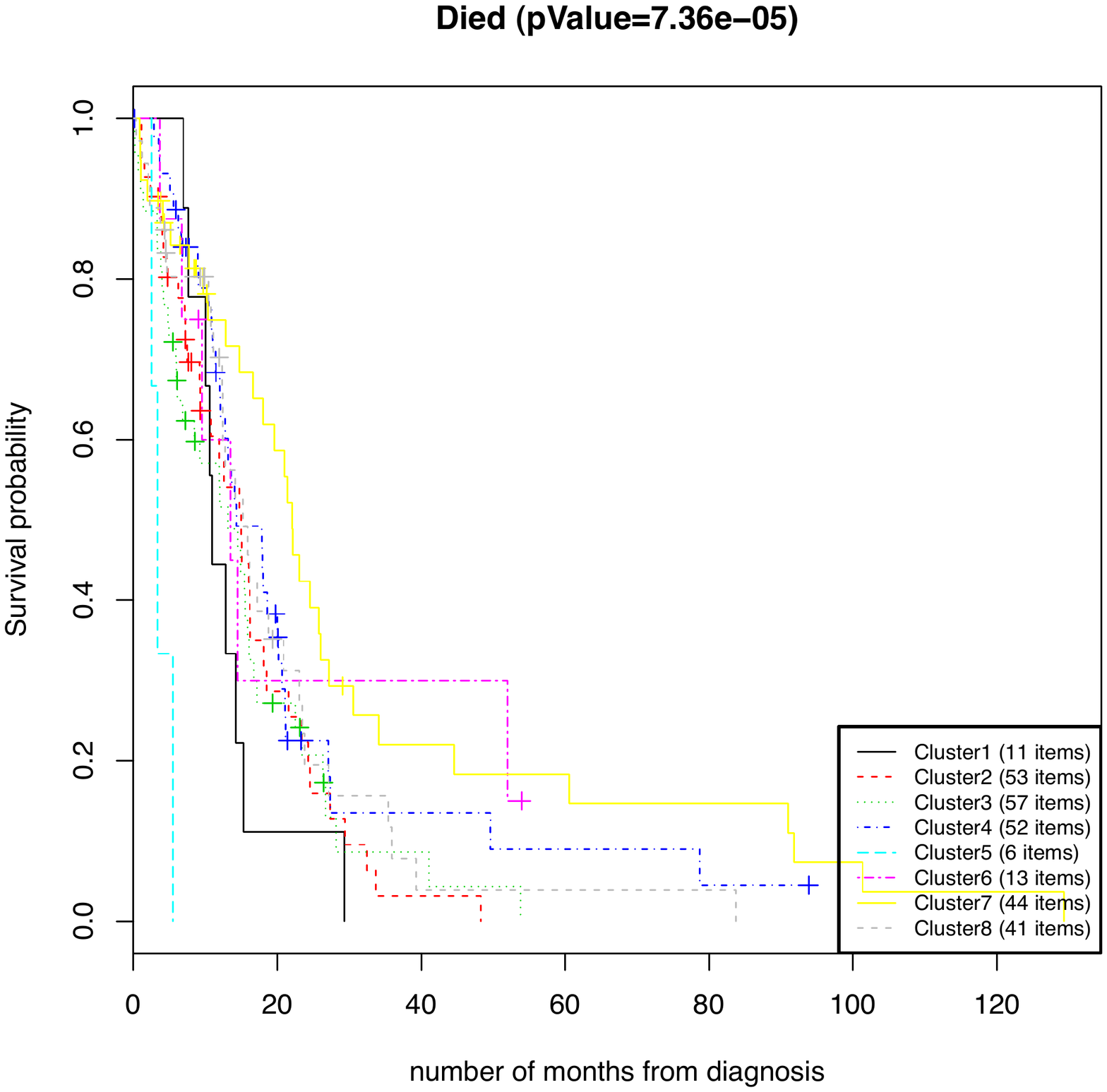}
\vspace{-15ex}
\caption{Kaplan-Meier survival curves for
  the gene expression disease subtypes.  The p-value is computed using a
  log-rank test.  When quoted in Table \ref{Table:qvalues}, a
  Bonferroni correction has been applied to account for multiple
  hypothesis tests.}
\label{fig:survival_death}
\end{figure*}

\begin{figure*}
\includegraphics[width=0.99\linewidth]{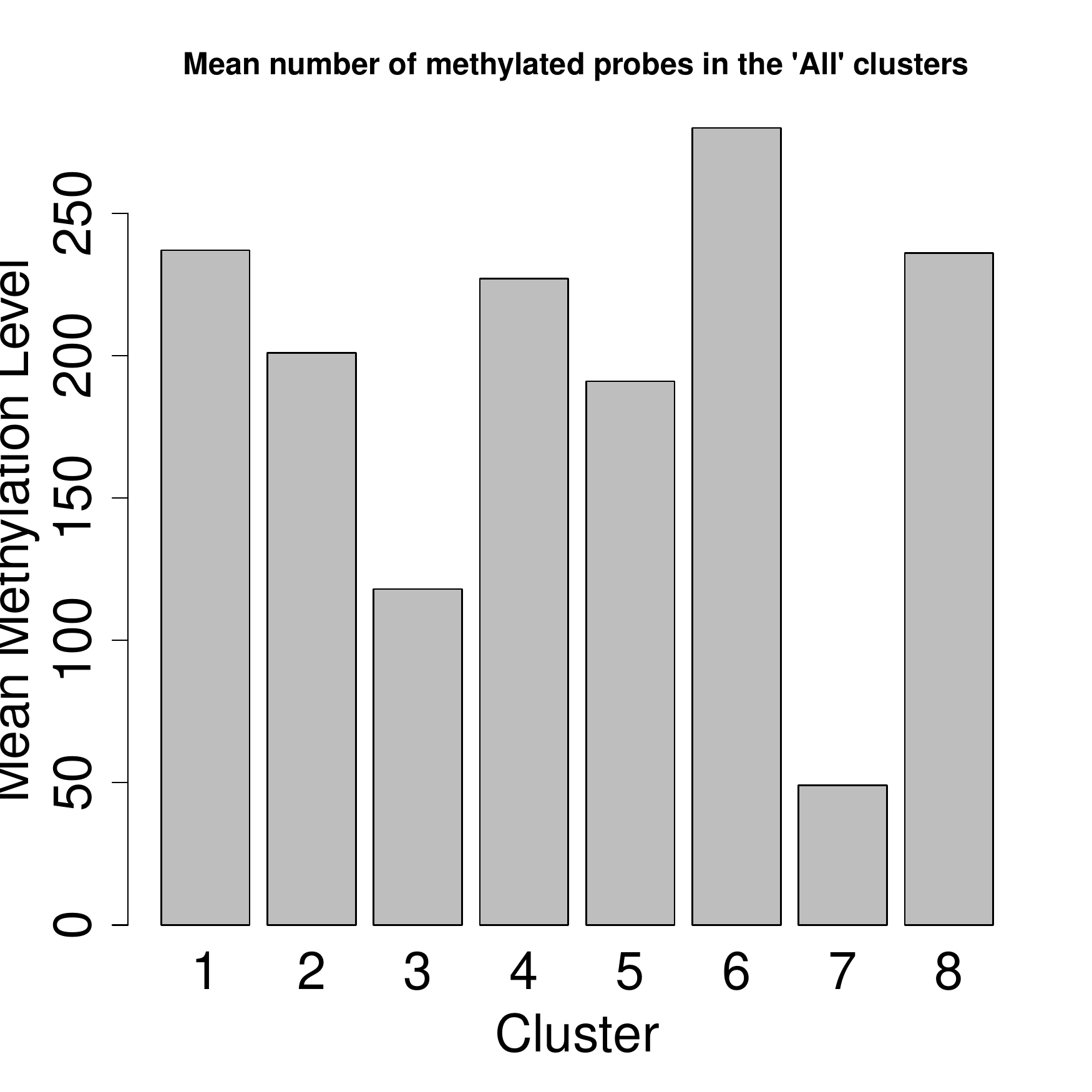}
\caption{The mean number of methylated sites in each consensus subtype
  (out of a possible 769).  A site is counted as methylated if it has
  been binarised to have unit value in the input methylation data in
  this analysis.  
}
\label{fig:methylationLevels}
\end{figure*}  

\begin{figure*}
\vspace{-30ex}
\centering
\includegraphics[width=1.2\linewidth]{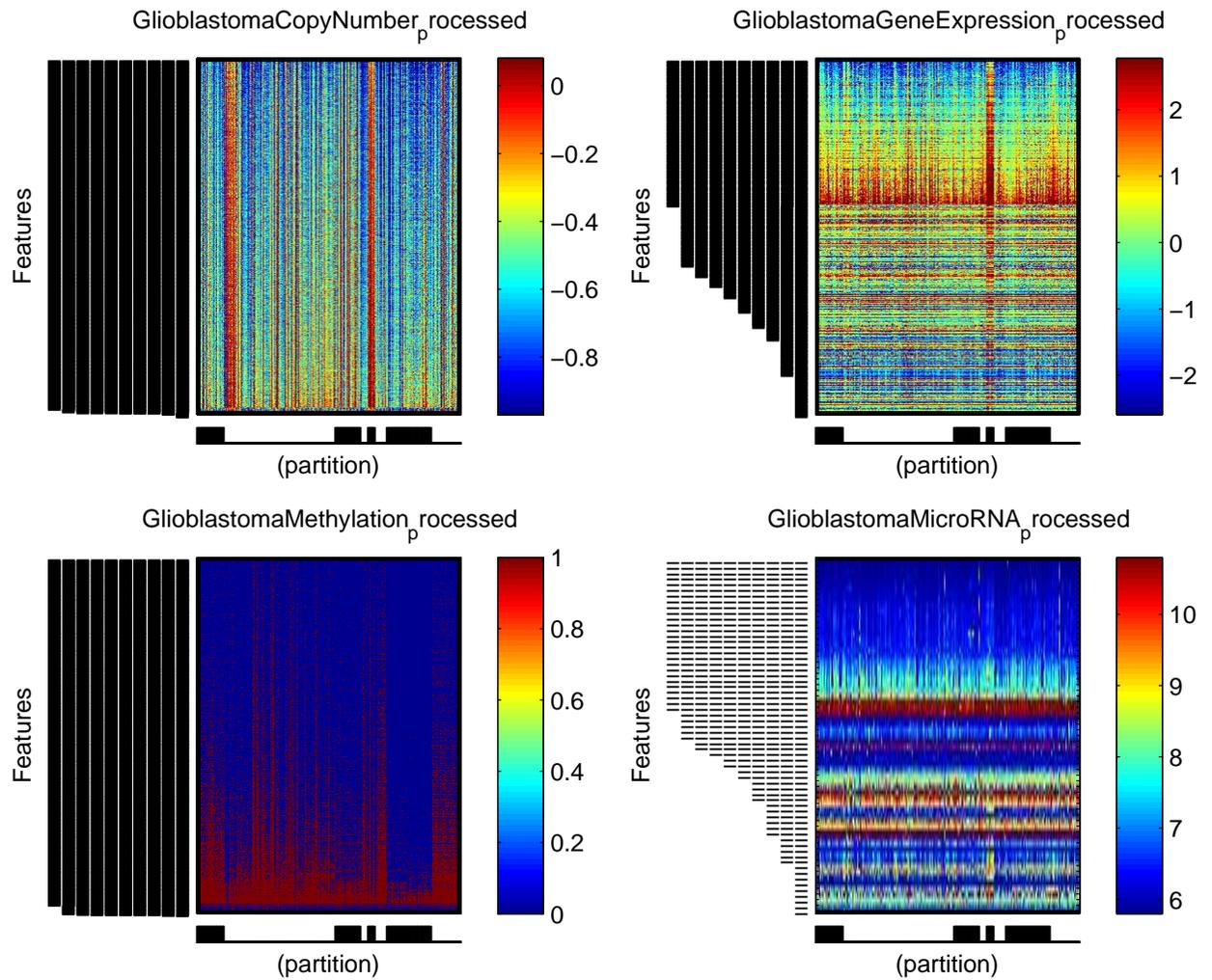}
\vspace{-30ex}
\caption{Plot of the 4 data types.  The x-axis gives the data items
  (patients/tumour samples), sorted by the consensus clustering
  partition.  The y-axis gives the selected features for each data
  type, with the dashed lines on the LHS indicating P(selected) for
  each feature.  We note that any feature with P(selected)$<$0.1 is
  excluded from this plot, and also that the outlying 5\% of pixels in
each data type are clipped for the purposes of plotting.}
\label{fig:allSortedData}
\end{figure*}  

\begin{figure*}
\centering
\includegraphics[width=1.1\linewidth]{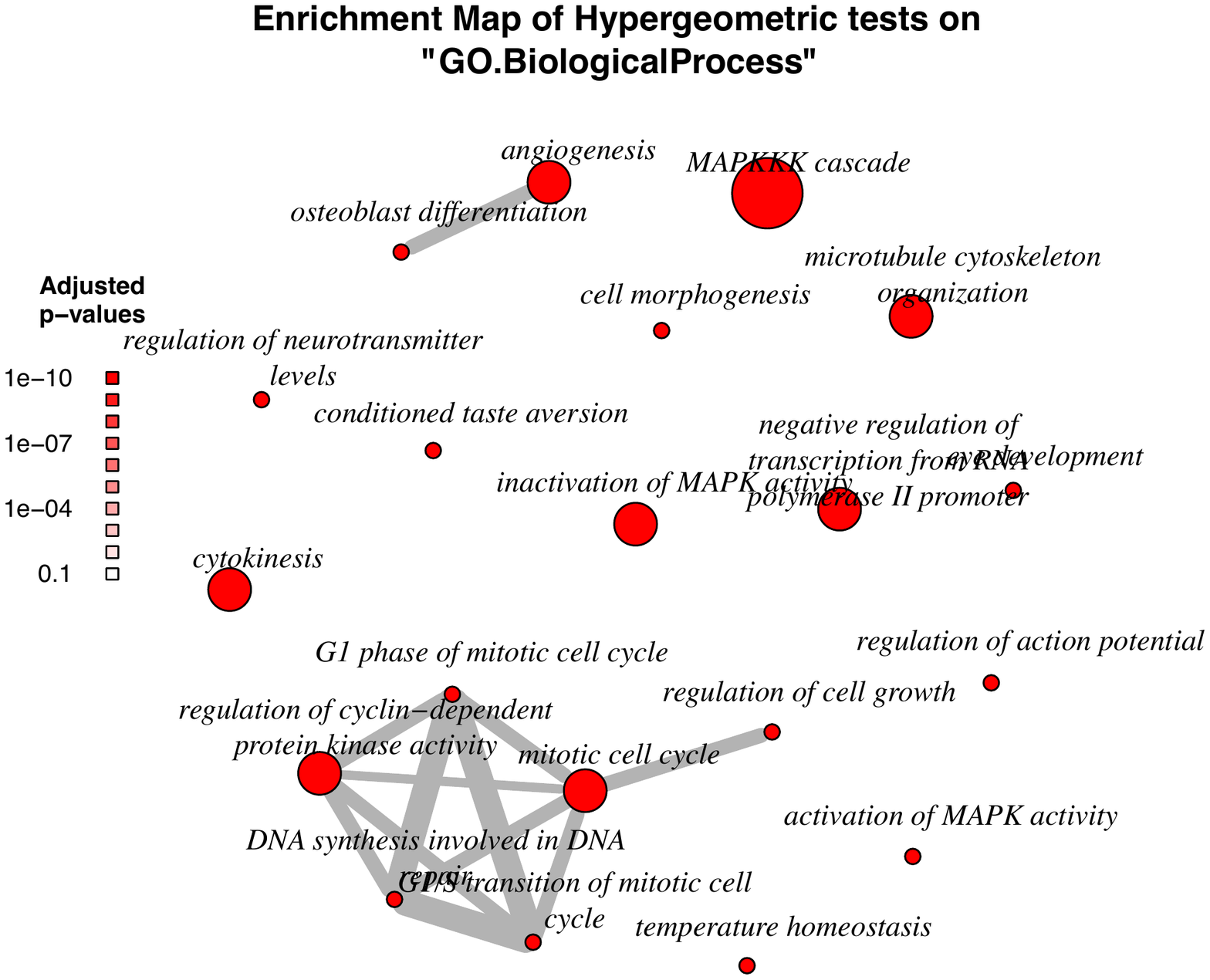}
\caption{
}
\label{fig:Enrichment_BP}
\end{figure*}  
\begin{figure*}
\centering
\includegraphics[width=1.1\linewidth]{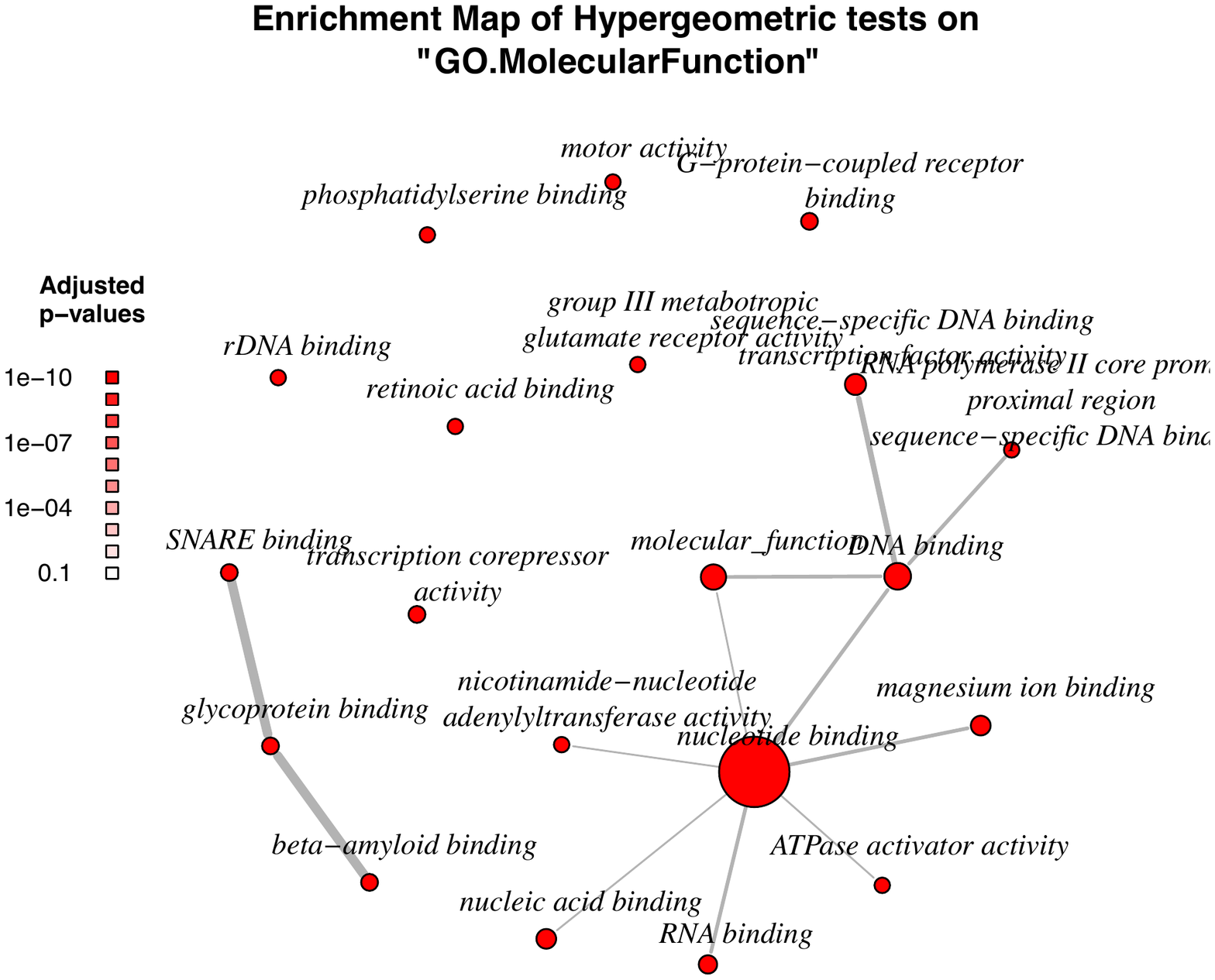}
\caption{
}
\label{fig:Enrichment_MF}
\end{figure*}  
\begin{figure*}
\centering
\includegraphics[width=1.1\linewidth]{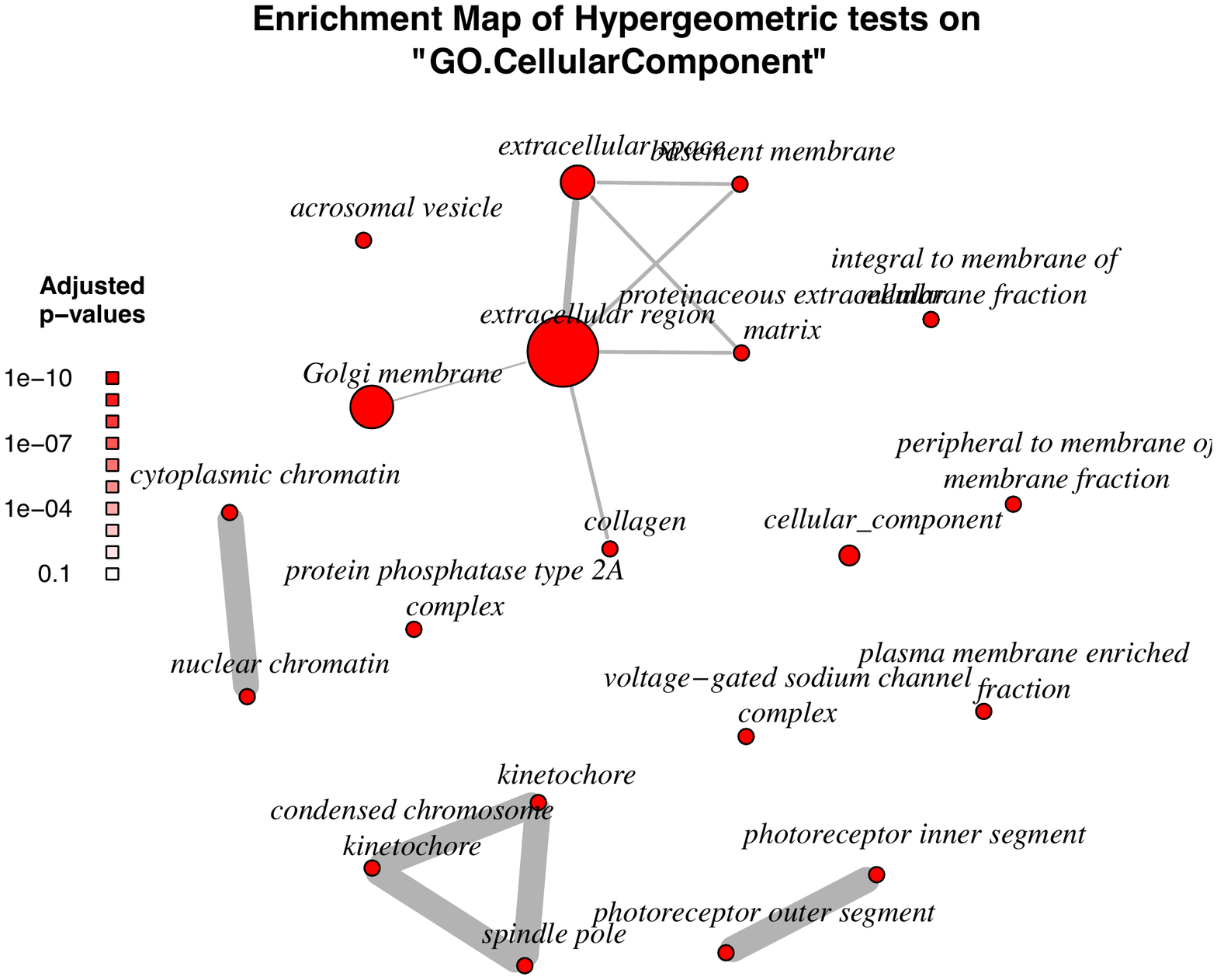}
\caption{
}
\label{fig:Enrichment_CC}
\end{figure*}  
\begin{figure*}
\centering
\includegraphics[width=1.1\linewidth]{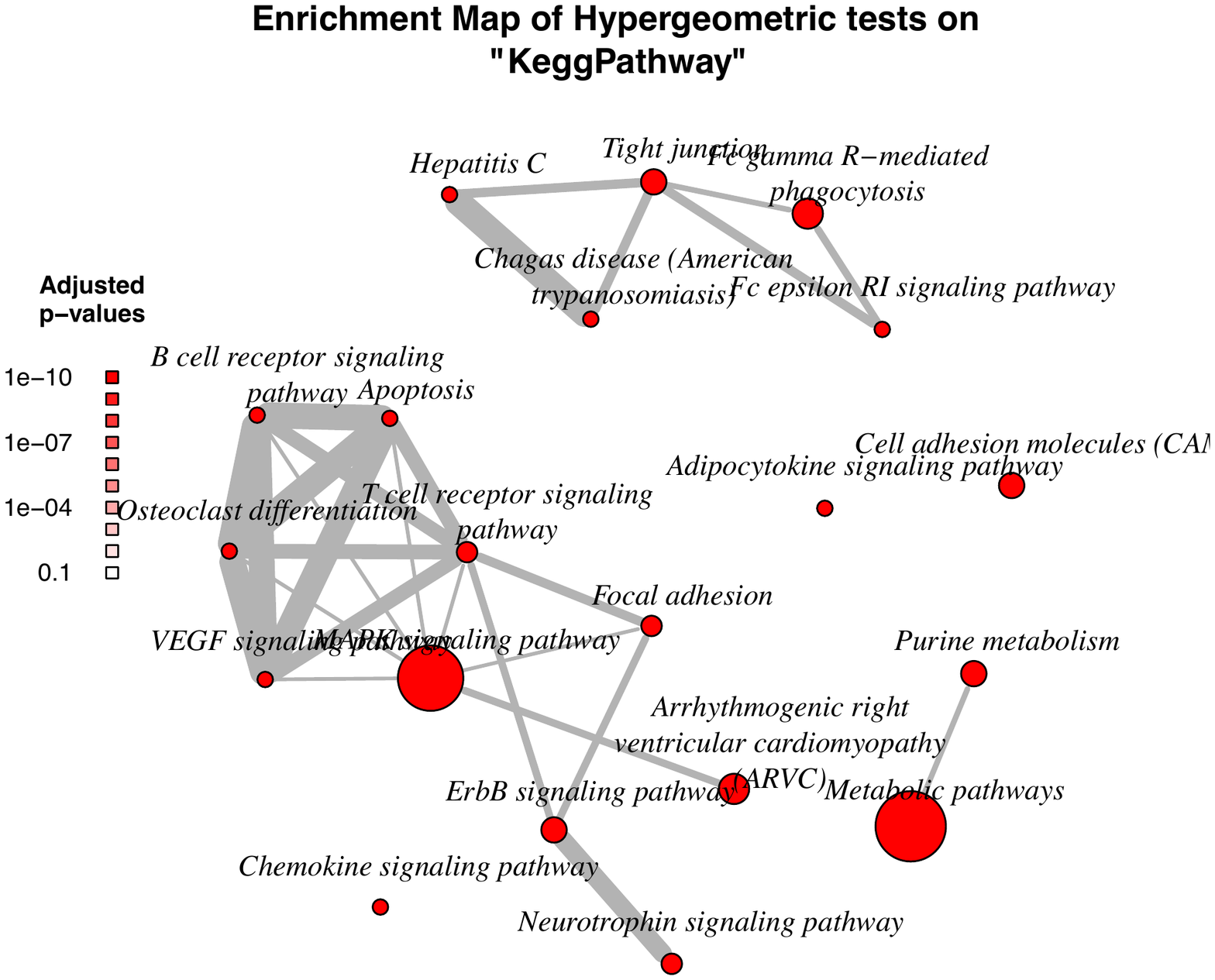}
\caption{
}
\label{fig:Enrichment_KEGG}
\end{figure*}


 


\bibliography{references.bib}
\bibliographystyle{icml2012}

\end{document}